\documentclass[aps,preprint]{revtex4}%
\usepackage{amsfonts}
\usepackage{amsmath}
\usepackage{amssymb}
\usepackage{graphicx}
\usepackage{pdfpages}
\usepackage{float}
\usepackage{color}
\usepackage{xcolor}
\usepackage{caption}
\usepackage{lineno,hyperref}
\usepackage{changes}%
\hypersetup{colorlinks =true, allcolors = blue}
\providecommand{\U}[1]{\protect\rule{.1in}{.1in}}

\begin{document}

\begin{center}
{\Large Greybody factors emitted by a regular black hole in a non-minimally coupled Einstein-Yang-Mills theory}

Ahmad Al-Badawi

Department of Physics, Al-Hussein Bin Talal University, P. O. Box: 20,
71111, Ma'an, Jordan

\bigskip E-mail: ahmadbadawi@ahu.edu.jo

{\Large Abstract}
\end{center}

In this paper, we study the greybody factors (GFs) for fermions with different spins and bosons in the regular black hole (BH) predicted by a non-minimal Einstein-Yang-Mills (EYM) theory. We investigate the effect of magnetic charge on effective potentials and GFs. For
this purpose, we consider the Dirac and Rarita-Schwinger, as well as
Klein-Gordon equations. First, we study the Dirac equation in curved spacetime for massive and massless spin-1/2 fermions. We then separate the Dirac equation into sets of radial
and angular equations. Using the analytical solution of the angular
equation, the Schr\"{o}dinger-like wave equations with potentials are derived by decoupling the radial wave equations via the tortoise coordinate. We also consider the Rarita-Schwinger equation for massless spin-3/2 fermions and derive the one-dimensional Schrödinger wave equation with gauge-invariant effective potential. For bosons, we study the Klein-Gordon equation in the regular
non-minimal EYM BH. Afterward, semi-analytic methods were used to calculate
the fermionic and bosonic GFs. Finally, we discuss
the graphical behavior of the obtained effective potentials and bounds on
the GFs. According to graphs, the GF is highly influenced by the potential's shape, which is determined by the parameterization of the model. This is in line with quantum theory.

\section{Introduction}

Black holes (BHs) are incredibly fascinating objects that Einstein's theory of general relativity (GR) predicted would exist. As a result of numerous astrophysical monitoring, it is now generally acknowledged that there are objects in the center of every galaxy that can be described by the spacetime geometry of BHs. Recent observation of the image of M87 and Sgr A*
supermassive BHs by Event Horizon Telescope collaboration [1-7] indicate
this conclusion. According to the data, Milky Way spiral galaxies and M87
elliptical galaxies have supermassive BHs masses of four million and six billion
solar masses, respectively. BHs are also now thought of as astrophysical laboratories for examining various expansions or adjusted theories of gravity, in addition to GR in the strong gravity regime. Furthermore, BHs have supplied important insights for comprehension the quantum nature of BH by connecting the laws of thermodynamics and
gravity. BHs are known to have entropy and to radiate energy called the
Hawking radiation. Thus, an observer positioned far away from BHs should be able to detect temperature [8-10]. Therefore, BHs are crucial in this regard for testing quantum gravity theory. The Hawking radiation must travel
through curved spacetime geometry before reaching an observer. The surrounding spacetime thus acts as a radiation barrier significantly modifies the black-body radiation spectrum observed by the asymptotic observer. Thus, the greybody factor (GF) is determined to calculate for the transmission amplitude of the BH's radiation. The rate of absorption
probability, which is another name for it, can be thought of as the
likelihood that a wave traveling from infinity will be absorbed by the BH.
The GF can be calculated using a variety of techniques , such as 
matching method [11-13], rigorous bound method [14,15], the WKB
approximation [16,17], and analytical methods for various spin fields
[18-23]. There is a large literature for topical reviews that the reader can
refer to when computing GFs [24-29].

Einstein-Yang-Mills (EYM) equations with the gauge group $SU(2)$ for any
event horizon have infinite BH solutions [30]. In general, several studies
have been conducted on the EYM theory of gravity and its BH solutions
[31-34]. The non-minimal EYM theory [35] was also expanded upon, and this
theory now admits new solutions for wormholes [36] and BHs [37,38], containing the non-minimal EYM BH spacetime we are focus on in this paper. Recently, the weak and strong deflection gravitational lensings by regular non-minimal EYM
BH was studied [39]. They found that the weak deflection lensing by such a
BH is the same as the one by a Reissner-Nordstr\"{o}m BH. While by using the
Newman-Janis algorithm on a spherically symmetric solution, the rotating version of the non-minimal EYM BH spacetime was found [40]. They have
studied the ergosurface and the BH shadow and discovered that the magnetic charge alters the BH shadow's size and shape.

 We concentrate on EYM BH because it is a nonsingular BH (regular). Regular BHs are possibly thought to be one of the solutions to the problem of the existence of singularities. Furthermore, the spacetime under consideration belongs to one of the classes of non-minimal field theories. These non-minimal theories give precise answers for electric and magnetic stars, wormholes, electric and regular magnetic BHs. On the other hand, studying the behavior of a spacetime under various types of perturbations, such as spinor and gauge fields, is an effective way to understand and analyze its properties [41,42]. Because the matter fields surrounding the BH are thought to be fermion fields, it becomes worthwhile to investigate the fluctuations as the fermion field.
The purpose of this work is to investigate spinorial wave equations, including Dirac, Rarita-Schwinger and Klein-Gordon equations, as well as GFs in a regular non-minimal EYM BH spacetime. We will be able to
derive the radial equations of the spinorial wave equations in the
background of EYM\ BH using the separation method and proper wave function
ansatzes. The radial equations will then be converted into Schr\"{o}%
dinger-like wave equations with their effective potentials using tortoise
coordinate. We will employ the common method of semi-analytical bounds to
obtain the GFs. In addition, we will graphically examine how the magnetic
charge parameter behaves in relation to the actual potential for Hawking
radiation absorption. It is of interest to consider radiating BHs in EYM
based on different interacting fields including the Yang-Mills (YM). As part
of this work, we aim to show how magnetic charge parameters affect the GFs emitted from EYM BH, and to contribute a theoretical study to the future studies on BH classification based on GF measurements.

The organization of the paper is as follows. Sect. II briefly reviewed the
spacetime of the regular non-minimal EYM BH. In Sect. III, we also reviewed the equations of motion related to the spinorial wave equations, namely the Dirac, Rarita-Schwinger and Klein-Gordon equations, around the EYM BH spacetime. Further, we obtain the corresponding effective potentials for the fields of fermions and bosons. In Sect. IV, using the effective potentials, the GFs are computed for massive and massless spin- 1/2 fermions, massless spin- 3/2 fermions and bosons, respectively. Further, we studied the calculation of the GFs of the BH's bounds and an analysis of their graphical behavior. In Sect. V, conclusion is given.

\section{Regular non-minimal EYM BH}

The action of the regular non-minimally EYM theory in four-dimensional spacetime is given by [37,38]%
\begin{equation}
S=\frac{1}{8\pi }\int d^{4}x\sqrt{-g}\left[ R+\frac{1}{2}F_{\mu \nu
}^{\left( a\right) }F^{\mu \nu {\left( a\right) }}+\frac{1}{2}R^{\alpha
\beta \mu \nu }F_{\alpha \beta }^{\left( a\right) }F_{\mu \nu }^{\left(
a\right) }\right] ,
\end{equation}%
where $g$ is the determinant of the metric tensor and $R$ is the Ricci
scalar. The Greek indices run from 0 to 3, while the Latin indices run from
1 to 3. Further, the symbol $F_{\mu \nu }^{\left( a\right) }$ denotes the YM tensor and is related to the YM potential $A_{\mu }^{\left( a\right) }$ 
 via the following equation 
\begin{equation}
F_{\mu \nu }^{\left( a\right) }=\nabla _{\mu }A_{\nu }^{\left( a\right)
}-\nabla _{\nu }A_{\mu }^{\left( a\right) }+f_{\left( b\right) \left(
c\right) }^{\left( a\right) }A_{\mu }^{\left( b\right) }A_{\nu }^{\left(
c\right) },
\end{equation}%
where $\nabla _{\mu }$ is the covariant derivative and the symbols $%
f_{\left( b\right) \left( c\right) }^{\left( a\right) }$\ denote the real
structure constants of the 3-parameters YM gauge group $SU\left( 2\right) $.
The tensor $R^{\alpha \beta \mu \nu }$ is given by [38]%
\begin{equation}
R^{\alpha \beta \mu \nu }=\frac{\xi _{1}}{2}\left( g^{\alpha \mu }g^{\beta \nu
}-g^{\alpha \nu }g^{\beta \mu }\right)
\end{equation}%
\begin{equation*}
+\frac{\xi _{2}}{2}\left( R^{\alpha \mu }g^{\beta \nu }-R^{\alpha \nu
}g^{\beta \mu }+R^{\beta \nu }g^{\alpha \mu }-R^{\beta \mu }g^{\alpha \nu
}\right) +\xi _{3}R^{\alpha \beta \mu \nu },
\end{equation*}%
where $R^{\alpha \beta }$ and $R^{\alpha \beta \mu \nu }$ are the Ricci and
Riemann tensors and \ $\xi _{i}$ $(i=1,2,3)$ are the non-minimal coupling
parameters between the YM field and the gravitational field. Assuming the gauge field to be described by the Wu-Yang ansatz and $%
\xi _{1}=\xi ,\xi _{2}=4\xi ,\xi _{3}=-6\xi $ along with $\xi >0$, a
regular, static, and spherically symmetric BH was found [37,38], 
\begin{equation}
ds^{2}=f\left( r\right) dt^{2}-f^{-1}(r)dr^{2}-r^{2}\left( d\theta ^{2}+\sin
^{2}\theta d\phi ^{2}\right)  \label{M1}
\end{equation}%
where 
\begin{equation}
f\left( r\right) =1+\left( \frac{r^{4}}{r^{4}+2\xi Q^{2}}\right) \left( 
\frac{Q^{2}}{r^{2}}-\frac{2M}{r}\right) ,
\end{equation}%
in which, $M$ is the BH mass and $Q$ is the magnetic charge. The
electromagnetic field four-potential of the regular non-minimal magnetic BH
reads%
\begin{equation}
A_{\nu }=\left( 0,0,0,Q\cos \theta \right) .
\end{equation}%
It is readily seen that for $Q=0$, the metric (\ref{M1}) reduces to the
Schwarzschild BH whereas for $\xi =0,$ it is Reissner-Nordstr\"{o}m BH
metric only with the magnetic charge instead of the electric charge. The number of the event horizons of the non-minimal EYM BH depends on the non-minimal parameter and the magnetic charge. There might be more than one horizon. In [43], it is demonstrated that as the coupling parameter is increased, the minimal outer horizon increases and reaches its maximum when $r_{h}=1.5M$ at $\xi \rightarrow \infty $. Figure 1 depicts the  variation of the metric function for different values of the magnetic charge parameter.
\begin{figure}
\centering
\includegraphics[scale=1]{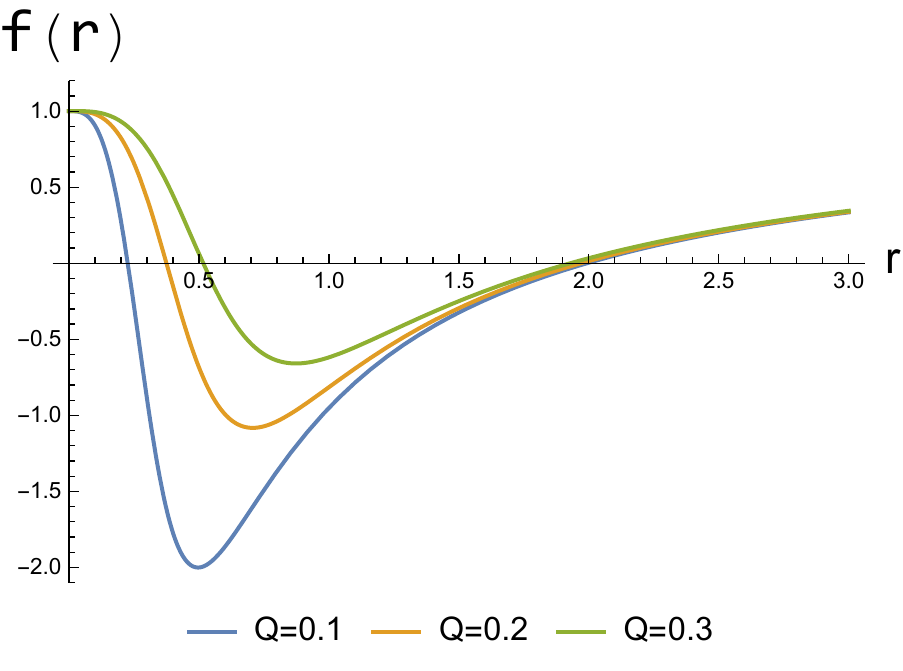} \caption{ Metric function of the EYM BH spacetime with for several values of $Q$. Here, we put the parameters as $M = 1 = \xi$}
\label{Figure1}%
\end{figure}

\section{\protect\bigskip\ Spinorial wave equations}

In this section, we will consider Dirac and Rarita-Schwinger as well as the Klein-Gordan equations in the EYM BH spacetime.

\subsection{Dirac equation}

We shall consider the vielbein formalism for the spin-1/2 fields in curved
spacetime, where the vielbein can be defined for our metric (\ref{M1}) as
\begin{equation}
e_{\widehat{\alpha }}^{\mu }=daig\left( \frac{1}{\sqrt{f}},\sqrt{f},\frac{1}{%
r},\frac{1}{r\sin \theta }\right) .
\end{equation}%
In curved spacetime, the Dirac equation for massive spin-1/2 particles is  
\begin{equation}
\gamma ^{\mu }\left[ \left( \partial _{\mu }+\Gamma _{\mu }\right) +m\right]
\Psi =0,  \label{q33}
\end{equation}%
where, $\gamma ^{\mu }$\ is the Dirac gamma matrix, $\Psi$ and $m$ are the Dirac field and its mass, and $\Gamma _{\mu }$ is
the spin connection. In terms of the Christoffel
symbol $\Gamma _{\mu \nu }^{\rho }$ the spin connection is written as
\begin{equation}
\Gamma _{\mu }=\frac{1}{8}e_{\widehat{\alpha }}^{\rho }\left( \partial _{\mu
}e_{\rho \widehat{\beta }}-\Gamma _{\mu \rho }^{\sigma }e_{\sigma \widehat{%
\beta }}\right) \left[ \gamma ^{\widehat{\alpha }},\gamma ^{\widehat{\beta }}%
\right] .
\end{equation}%
Here, the Dirac gamma matrices, $\gamma ^{\widehat{\mu }}$ are represented in terms of
the Pauli spin matrices $\sigma ^{i}\left( i=0,1,2,3\right) .$ With this
setting, the Dirac equation (\ref{q33}) can be rewritten as%
\begin{equation}
\left[ \frac{1}{\sqrt{f}}\left( \gamma ^{\widehat{0}}\partial _{t}+\frac{f^{\prime }}{4
}\gamma ^{\widehat{1}}\right) +\sqrt{f}\gamma ^{\widehat{1}}\partial _{r}+\frac{1}{r}\gamma ^{\widehat{2}}\partial_{\theta }+\frac{\sqrt{f}}{2r}\gamma^{\widehat{1}}\right.  \label{q34}
\end{equation}%
\begin{equation*}
\left. +\frac{1}{r\sin \theta }\gamma ^{\widehat{3}}\partial _{\phi }+\frac{\sqrt{f}}{%
r}\gamma ^{\widehat{1}}+\frac{\cot \theta }{2r}\gamma ^{\widehat{2}} + m \right] \Psi =0,
\end{equation*}%
where prime denotes the derivative with respect to r. To solve Eq. (\ref{q34}%
) we use the separation variable method as%
\begin{equation}
\Psi \left( t,r,\theta ,\phi \right) =\left( 
\begin{array}{c}
iAe^{-i\omega t} \\ 
Be^{-i\omega t}%
\end{array}%
\right) \otimes \Theta \left( \theta ,\phi \right) ,  \label{q35}
\end{equation}%
where $A=A\left( r\right) $, $B=B\left( r\right) $ and $\omega $ is the
angular frequency of the solution. Substituting Eq. (\ref{q35})\ into Eq. (%
\ref{q34}), then the radial part of the Dirac equation is obtained as
follows:%
\begin{equation}
\left[ \left( f\partial _{r}+\frac{f^{\prime }}{4}+\frac{f}{r}\right) \sigma
^{2}+\frac{i\lambda \sqrt{f}}{r}\sigma ^{1}\right] \left( 
\begin{array}{c}
iA \\ 
B%
\end{array}%
\right) =-\left[ \omega \sigma ^{3}+m\sqrt{f}\right] \left( 
\begin{array}{c}
iA \\ 
B%
\end{array}%
\right) ,  \label{q36}
\end{equation}%
where $\lambda =l+1=\pm 1,\pm 2,\pm 3,...$ are the eigenvalues for the angular part that satisfy the following angular equation%
\begin{equation}
\left[ \sigma ^{1}\partial _{\theta }-\frac{\sigma ^{2}}{\sin \theta }%
\partial _{\phi }+\frac{\cot \theta }{2}\sigma ^{1}\right] \Theta =i\lambda
\Theta .
\end{equation}%
To decouple the redial equations (\ref{q36}), we define new radial functions 
$F\left( r\right) $ and $G\left( r\right) ,$ (for more details see [44]) as
follows%
\begin{equation}
\frac{d^{2}F}{dr_{\ast }^{2}}+\left( \omega ^{2}-V_{+1/2}\right) F=0,
\end{equation}%
\begin{equation}
\frac{d^{2}G}{dr_{\ast }^{2}}+\left( \omega ^{2}-V_{-1/2}\right) G=0,
\end{equation}%
where $r_{\ast }$ is the tortoise coordinate defined by 
\begin{equation}
\frac{dr_{\ast }}{dr}=\frac{2\omega \left( \lambda ^{2}+m^{2}r^{2}\right) }{%
2\omega \left( \lambda ^{2}+m^{2}r^{2}\right) +fm\lambda }f.
\end{equation}%
Thus, the effective potentials of the massive fermionic waves having spin- 1/2 and moving in the EYM\ BH geometry are, 
\begin{equation}
V_{\pm 1/2}=\pm \frac{dW}{dr_{\ast }}+W^{2},  \label{p9}
\end{equation}%
where%
\begin{equation}
W=\frac{\left( \sqrt{f}/r\right) \sqrt{\lambda ^{2}+m^{2}r^{2}}}{1+\left(
f/2\omega \right) \left( \lambda m/\left( \lambda ^{2}+m^{2}r^{2}\right)
\right) }.  \label{p10}
\end{equation}%
We note that the potential depends mainly on the metric function and in addition to the mass and the energy of the Dirac field. The explicit forms of the effective potentials for massive spin- 1/2 fermions are written as
\begin{equation*}
V_{\pm 1/2}=\frac{\left( \lambda ^{2}+m^{2}r^{2}\right) ^{3/2}\sqrt{f}}{D^{2}%
}\left[ \frac{\sqrt{f}}{r^{2}}\left( \lambda ^{2}+m^{2}r^{2}\right)
^{3/2}\pm \left( \frac{f^{\prime }}{2r}-\frac{f}{r^{2}}\right)\left( \lambda ^{2}+m^{2}r^{2}\right) \pm 3m^{2}f%
\right]
\end{equation*}%
\begin{equation}
\mp \frac{\left( \lambda ^{2}+m^{2}r^{2}\right) ^{5/2}f^{3/2}}{rD^{3}}\left[
2m^{2}r+\frac{\lambda m}{2\omega}f^{\prime} \right] ,  \label{p11}
\end{equation}%
where%
\begin{equation*}
D=\left( \lambda ^{2}+m^{2}r^{2}+ \frac{\lambda m}{2\omega} f\right) .
\end{equation*}%
While the effective potential for massless spin- 1/2 fermions Dirac fields propagating in EYM BH spacetime is obtained by setting $m=0$ in (\ref{p11}) namely
\begin{equation}
V_{\pm 1/2}=\frac{\lambda }{r^{2}}\left( \lambda f\pm \frac{rf^{\prime }}{2%
\sqrt{f}}\mp f^{3/2}\right) .  \label{p12}
\end{equation}%

\subsection{Rarita-Schwinger equation}

In this part, we will consider the Rarita-Schwinger equation to represent the massless spin- 3/2 field, 
\begin{equation}
\gamma ^{\mu \nu \alpha }\widetilde{\emph{D}}_{\nu }\psi _{\alpha }=0
\label{Rs 1}
\end{equation}%
where $\widetilde{\emph{D}}_{\nu }$ is the supercovariant derivative, $\psi
_{\alpha }$ indicates the spin-3/2 field and $\gamma ^{\mu \nu \alpha }$ is
the anti symmetric of Dirac gamma matrices given by 
\begin{equation}
\gamma ^{\mu \nu \alpha }=\gamma ^{\lbrack \mu }\gamma ^{\nu }\gamma
^{\alpha ]}=\gamma ^{\mu }\gamma ^{\nu }\gamma ^{\alpha }-\gamma ^{\mu
}g^{\nu \alpha }+\gamma ^{\nu }g^{\mu \alpha }-\gamma ^{\alpha }g^{\mu \nu }.
\end{equation}%
The supercovariant derivative for the spin-3/2 field in our BH spacetime can
be written as%
\begin{equation}
\widetilde{\emph{D}}_{\nu }=\nabla _{\nu }+\frac{1}{4}\gamma _{\alpha
}F_{\nu }^{\alpha }+\frac{i}{8}\gamma _{\nu \alpha \mu }F^{\alpha \mu }.
\end{equation}%
Our attention is now on the non-TT eigenfunctions [45], so we shall provide
the radial wave function and the temporal wave function in the following
order:%
\begin{equation}
\psi _{r}=\phi _{r}\otimes \overline{\psi }_{\left( \lambda \right) }\text{
and }\psi _{t}=\phi _{t}\otimes \overline{\psi }_{\left( \lambda \right) },
\end{equation}%
where $\overline{\psi }_{\left( \lambda \right) }$ is an eigenspinor with
eigenvalue $i\overline{\lambda }=i\left( j+1/2\right) ,$where $%
j=3/2,5/2,7/2,...$. The angular wave function is written as%
\begin{equation}
\psi _{\theta i}=\phi _{\theta }^{\left( 1\right) }\otimes \overline{\nabla }%
_{\theta i}\overline{\psi }_{\left( \lambda \right) }+\phi _{\theta
}^{\left( 2\right) }\otimes \overline{\gamma }_{\theta i}\overline{\psi }%
_{\left( \lambda \right) },
\end{equation}%
where $\phi _{\theta }^{\left( 1\right) },\phi _{\theta }^{\left( 2\right) }$
are functions of $r$ and $t.$ We will consider the cases $\mu =t,r,\theta
_{i}$ in Eq. (\ref{Rs 1}) respectively to obtain the equations of motion [46] subsequently,
\begin{equation}
0=-\left[ i\overline{\lambda }+\sqrt{f} i\sigma ^{3}\right] \phi _{r}+\left[ i%
\overline{\lambda }\partial _{r}-\frac{1}{2r\sqrt{f}}\left( i\sigma
^{3}\right) +\frac{i\overline{\lambda }}{2r}\right] \phi _{\theta}^{\left(
1\right) }  \label{R10}
\end{equation}%
\begin{equation*}
+\left[ 2\partial _{r}+\frac{i\overline{\lambda }}{r\sqrt{f}}i\sigma ^{3}+%
\frac{1}{r}\right] \phi _{\theta}^{\left( 2\right) },
\end{equation*}%
\begin{equation}
0=\left[ -\frac{i\overline{\lambda }}{\sqrt{f}}\partial _{t}+\frac{i%
\overline{\lambda }f^{\prime }}{4\sqrt{f}}\sigma ^{1}-\frac{1}{2r}\sigma
^{2}+\frac{i\overline{\lambda }\sqrt{f}}{2r}\sigma ^{1}\right] \phi _{\theta
}^{\left( 1\right) }  \label{R12}
\end{equation}%
\begin{equation*}
+\left[ -\frac{2}{\sqrt{f}}\partial _{t}+\frac{f^{\prime }}{2\sqrt{f}}\sigma
^{1}+\frac{i\overline{\lambda }}{r}\sigma ^{2}+\frac{\sqrt{f}}{r}\sigma ^{1}%
\right] \phi _{\theta }^{\left( 2\right) },
\end{equation*}%
\begin{equation}
0=\left[ \frac{1}{r\sqrt{f}}i\sigma ^{3}\partial _{t}+\frac{\sqrt{f}}{r}%
\sigma ^{2}\partial _{r}+\frac{f^{\prime }}{4r\sqrt{f}}\sigma ^{2}\right]
\phi ^{\left( 1\right) }-\left[ \frac{\sqrt{f}}{r}\sigma ^{2}\right] \phi
_{r},  \label{R14}
\end{equation}%
where $\sigma ^{i}\left( i=1,2,3\right) $ are the Pauli matrices. The gauge
invariant variable can be obtained by using the Weyl gauge $\left( \phi
_{t}=0\right) $ and the same arguments as in [46],%
\begin{equation}
\Phi =-\left( \frac{\sqrt{f}}{2}i\sigma ^{3}\right) \phi _{\theta }^{\left(
1\right) }+\phi _{\theta }^{\left( 2\right) }.
\end{equation}%
To obtain the effective potentials we will use the gauge-invariant variable $%
\Phi $, hence, the equations (\ref{R10}-\ref{R14}) become%
\begin{equation}
\left( 2\partial _{r}+\frac{i\overline{\lambda }}{r\sqrt{f}}\left( i\sigma
^{3}\right) +\frac{1}{r}\right) \Phi +\left( i\overline{\lambda }+\sqrt{f}%
i\sigma ^{3}\right) \partial _{r}\phi _{\theta} ^{\left( 1\right) }=\left( i\overline{%
\lambda }+\sqrt{f}i\sigma ^{3}\right) \phi _{r},  \label{R24}
\end{equation}%
\begin{equation}
\left( -\frac{2}{\sqrt{f}}\partial _{t}+\frac{\sqrt{f}}{r}\sigma ^{1}+\frac{%
f^{\prime }}{2\sqrt{f}}\sigma ^{1}+\frac{i\overline{\lambda }}{r}\sigma
^{2}\right) \Phi   \label{R25}
\end{equation}%
\begin{equation*}
+\left( -\left( \frac{i\overline{\lambda }}{\sqrt{f}}+i\sigma ^{3}\right)
\partial _{t}+\frac{i\overline{\lambda }f^{\prime }}{4\sqrt{f}}\sigma ^{1}-%
\frac{f^{\prime }}{4}\sigma ^{2}\right) \phi _{\theta }^{\left( 1\right) }=0,
\end{equation*}%
\begin{equation}
\left( \frac{1}{\sqrt{f}}\left( i\sigma ^{3}\right) \partial _{t}+\sqrt{f}%
\sigma ^{2}\partial _{r}+\frac{f^{\prime }}{4\sqrt{f}}\sigma ^{2}\right)
\phi _{\theta }^{\left( 1\right) }-\sqrt{f}\sigma ^{2}\phi _{r}=0
\label{R26}
\end{equation}%
The gauge-invariant equation of motion can be written in terms of only $\Phi 
$ by using the equations (\ref{R24}-\ref{R26}), thus
\begin{equation}
0=\left( \sqrt{f}+\overline{\lambda }\sigma ^{3}\right) \left[ -2\sigma
^{1}\partial _{t}+\frac{1}{2}f^{\prime }+\frac{\sqrt{f}}{r}\left( \sqrt{f}-%
\overline{\lambda }\sigma ^{3}\right) \right] \Phi   \label{R27}
\end{equation}%
\begin{equation*}
-\left( \sqrt{f}-\overline{\lambda }\sigma ^{3}\right) \left[ 2f\partial
_{r}+\frac{\sqrt{f}}{r}\left( \sqrt{f}-\overline{\lambda }\sigma ^{3}\right) %
\right] \Phi .
\end{equation*}%
Or it can be written as \ 
\begin{equation}
\Phi =\left( 
\begin{array}{c}
\phi _{1}e^{-i\omega t} \\ 
\phi _{2}e^{-i\omega t}%
\end{array}%
\right) ,
\end{equation}%
where $\phi _{1}$ and $\phi _{2}$ are purely radially dependent terms.
Hence, Eq. (\ref{R27}) becomes 
\begin{equation}
\left[ \left( \frac{\sqrt{f}-\overline{\lambda }}{\sqrt{f}+\overline{\lambda 
}}\right) f\partial _{r}-\frac{f^{\prime }}{4}-\left( \frac{\sqrt{f}-%
\overline{\lambda }}{\sqrt{f}+\overline{\lambda }}\right) \frac{\sqrt{f}}{r}%
\overline{\lambda }\right] \phi _{1}=i\omega \phi _{2}  \label{R29}
\end{equation}%
\begin{equation}
\left[ \left( \frac{\sqrt{f}+\overline{\lambda }}{\sqrt{f}-\overline{\lambda 
}}\right) f\partial _{r}-\frac{f^{\prime }}{4}-\left( \frac{\sqrt{f}+%
\overline{\lambda }}{\sqrt{f}-\overline{\lambda }}\right) \frac{\sqrt{f}}{r}%
\overline{\lambda }\right] \phi _{2}=i\omega \phi _{1}  \label{R30}
\end{equation}%
The above equations (\ref{R29}-\ref{R30}) can be simplified by redefining 
\begin{equation}
\phi _{1}=\left(\frac{\sqrt{f}+\overline{\lambda }}{
f^{1/4}}\right) \overline{\phi }_{1},\qquad \phi _{2}=\left(\frac{\sqrt{f}-\overline{\lambda }}{
f^{1/4}}\right) \overline{\phi }_{2}
\end{equation}%
Therefore, Eq. (\ref{R29}) and Eq. (\ref{R30}) are then cast into 
\begin{equation}
\left( \frac{d}{dr_{\ast }}-W\right) \overline{\phi }_{1}=i\omega \overline{%
\phi }_{2},
\end{equation}%
\begin{equation}
\left( \frac{d}{dr_{\ast }}+W\right) \overline{\phi }_{2}=i\omega \overline{%
\phi }_{1},
\end{equation}%
where the tortoise coordinate $r_{\ast }$ is defined as $\frac{d}{dr_{\ast }}%
=f\frac{d}{dr}$ and $W$ is obtained from 
\begin{equation}
W=\frac{\left\vert \overline{\lambda }\right\vert \sqrt{f}}{r}\left( \frac{%
\left\vert \overline{\lambda }\right\vert ^{2}-1}{\left\vert \overline{%
\lambda }\right\vert ^{2}-f}\right) .
\end{equation}%
Finally, we are able to describe our particles using the Schr\"{o}dinger-like wave equation as 
\begin{equation}
\left( -\frac{d^{2}}{dr_{\ast }^{2}}+V_{1}\right) \overline{\phi }_{1}=\omega
^{2}\overline{\phi }_{1},
\end{equation}%
\begin{equation}
\left( -\frac{d^{2}}{dr_{\ast }^{2}}+V_{2}\right) \overline{\phi }%
_{2}=\omega ^{2}\overline{\phi }_{2},
\end{equation}%
where the effective potentials $V_{1,2}$ can be determined as 
\begin{equation}
V_{1,2}=\pm f\frac{dW}{dr}+W^{2}.
\end{equation}%
Thus, the explicit forms of the effective potentials of EYM BH spacetime for massless spin-3/2 fermions are written as%
\begin{equation}
V_{1,2}=f\frac{\overline{\lambda }\left( 1-\overline{\lambda }^{2}\right) }{%
r\left( f-\overline{\lambda }^{2}\right) }\left[ \frac{\overline{\lambda }%
\left( 1-\overline{\lambda }^{2}\right) }{r\left( f-\overline{\lambda }%
^{2}\right) }\pm \left( \frac{rf^{\prime }-2f}{2r\sqrt{f}}\right) \mp \frac{%
\sqrt{f}f^{\prime }}{\left( f-\overline{\lambda }^{2}\right) }\right] .
\label{p32}
\end{equation}

\subsection{\protect\bigskip Klein-Gordon equation}

We will study in this part the the Klein-Gordon equation to represents the massless scalar
field, 
\begin{equation}
\frac{1}{\sqrt{-g}}\partial _{\mu }\sqrt{-g}g^{\mu \nu }\partial _{\nu
}U(t,r,\theta ,\phi )=0,  \label{is1}
\end{equation}%
where $\sqrt{-g}=r^{2}\sin \theta $ for metric (\ref{M1}). Assuming a separable solution 
\begin{equation}
U=R\left( r\right) Y_{m}^{l}\left(
\theta ,\phi \right) \exp \left( -i\omega t\right),
\end{equation} 
where $\omega$ denotes the frequency of the wave, $m$ is the azimuthal quantum number and $Y_{m}^{l}$ are the usual spherical harmonics. Thus, Eq. (\ref{is1}) can be
decouple into radial and angular sets. Then, the radial one can be transformed to a one dimensional Schr\"{o}dinger-like wave equation as follows, 
\begin{equation}
\frac{d^{2}U}{dr_{\ast }^{2}}+\left( \omega ^{2}-V_{eff}\right) U=0,
\label{s1}
\end{equation}%
where $r_{\ast }$ is the tortoise coordinate: $\frac{dr_{\ast }}{dr}=\frac{1%
}{f},$ and $V_{eff}$ is the effective potential given by%
\begin{equation}
V_{eff}=f\left( \frac{l(l+1)}{r^{2}}+\frac{f^{\prime }}{r}\right) .
\label{v1}
\end{equation}%
where $l$ is the angular quantum number.

\section{Greybody factors of EYM BH}

GFs are a measure of how much the spectrum of radiation emitted by a BH deviates from that of a perfect black body. The effective potentials will be used in this section to compute the GFs of fermions and bosons using general semi-analytic bounds. Hence, we are able to qualitatively analyze the results. Consequently, it is possible to determine
how the potential affects the GF. The transmission probability $%
\sigma _{l}\left( w\right) $ is given by 
\begin{equation}
\sigma _{l}\left( w\right) \geq \sec h^{2}\left( \int_{-\infty }^{+\infty
}\wp dr_{\ast }\right) ,  \label{is8}
\end{equation}

where $r_{\ast }$ is the tortoise coordinate and 
\begin{equation}
\wp =\frac{1}{2h}\sqrt{\left( \frac{dh\left( r_{\ast }\right) }{dr_{\ast }}%
\right) ^{2}+(w^{2}-V_{eff}-h^{2}\left( r_{\ast }\right) )^{2}}.  \label{is9}
\end{equation}

where $h(r_{\ast })$ is a positive function satisfying $h\left( -\infty
\right) =h\left( -\infty \right) =w$. For more details, one can see [47]. We
select $h=w$. Therefore, Eq. (\ref{is8}) 
\begin{equation}
\sigma _{l}\left( w\right) \geq \sec h^{2}\left( \int_{r_{h}}^{+\infty }%
\frac{V_{eff}}{2w}dr_{\ast }\right) .  \label{gb1}
\end{equation}%
In this process, the metric function plays a significant role in determining
the relationship between the GFs and the effective potential. Our GFs
calculations will be carried out in four cases since we have fermions with
different spins and bosons. When computing GFs, we usually focus on the study of one potential for each field.

\subsection{Spin- 1/2 fermions emission}

\subsubsection{Massive case}

Substituting the effective potential (\ref{p9}) derived from Dirac equation
into Eq. (\ref{gb1}), we obtain
\begin{equation}
\sigma _{l}\left( w\right) \geq \sec h^{2}\left[ \frac{1}{2w}\left(
\int_{r_{h}}^{+\infty }\left( \frac{dW}{dr_{\ast }}\right) dr_{\ast
}+\int_{r_{h}}^{+\infty }W^{2}dr_{\ast }\right) \right] .  \label{in10}
\end{equation}%
We will discuss the first and second integrals in Eq (\ref{in10}) separately as follows. For the first integral we have%
\begin{equation}
\int_{r_{h}}^{+\infty }\left( \frac{dW}{dr_{\ast }}\right) dr_{\ast
}=1,\label{in32}
\end{equation}%
since $\lim_{r\rightarrow \infty }W=\mu$ and $W\left( r_{h}\right) =0$ (the function $W$ is proportional to $f$ and $f$ vanishes at the horizons) . However, the second integral can be written as 
\begin{equation}
\int_{r_{h}}^{+\infty }\left( \frac{2\omega \left[ \lambda mf+2\omega \left(
\lambda ^{2}+m^{2}r^{2}\right) \right] }{m^{2}r^{2}\left( \lambda
^{2}+m^{2}r^{2}\right) f^{2}}\right) dr,  \label{in11}
\end{equation}%
\begin{figure}
    \centering
{{\includegraphics[width=7.5cm]{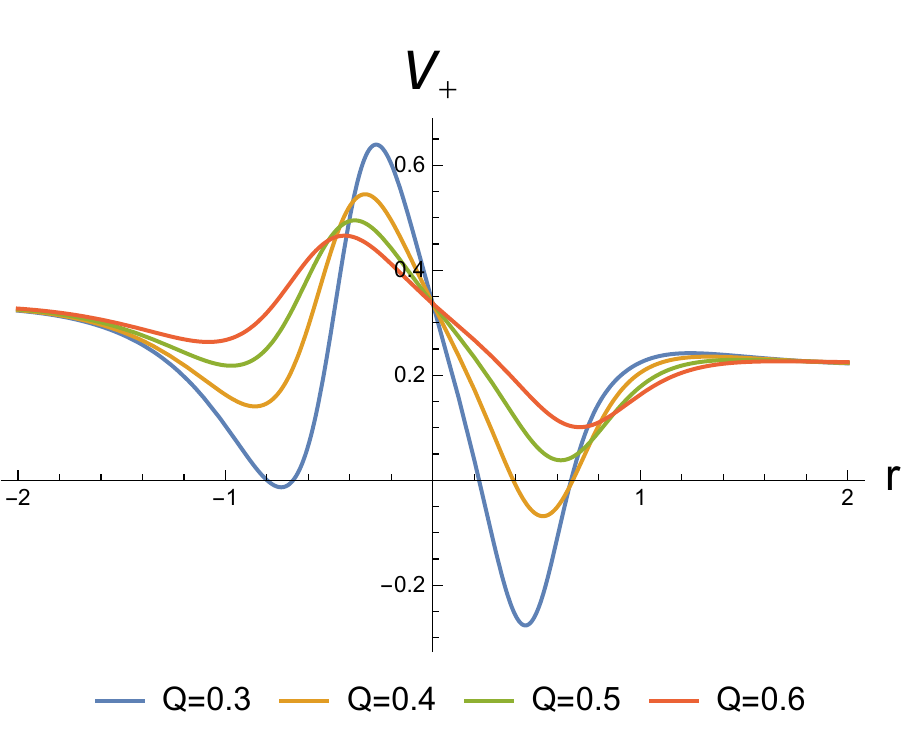} }}\qquad
    {{\includegraphics[width=7.5cm]{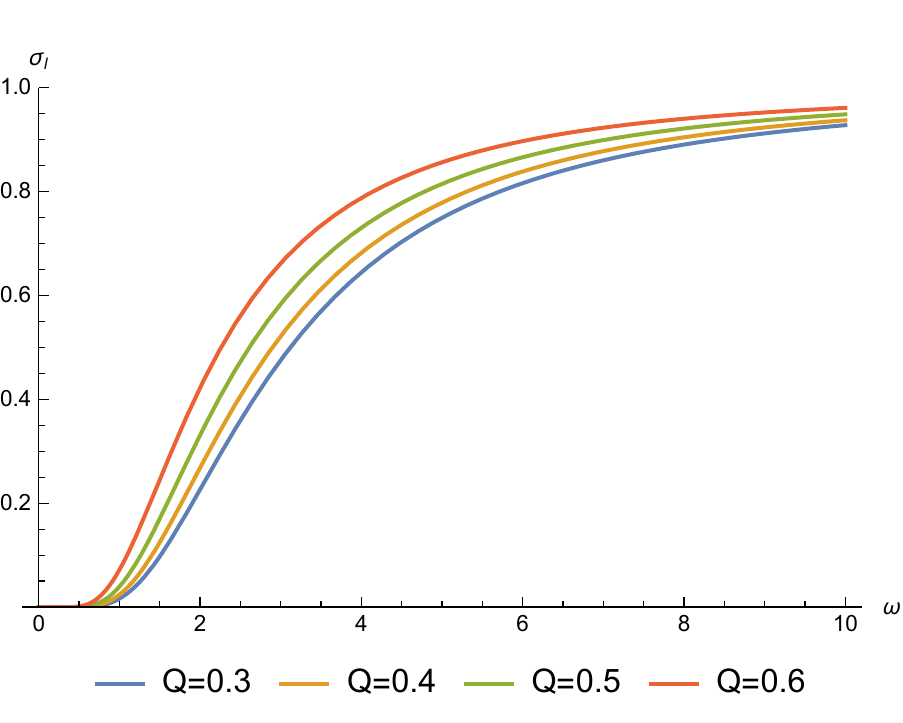}}}
    \caption{The potentials (\ref{p11})
for massive spin- 1/2 fermions for several values of $Q$ (left panel). The corresponding GF (right panel).
Here, we put the parameters as $\lambda =  1 = m = \omega = M = \xi$.}%
    \label{fig1}%
\end{figure}
In order to overcome the difficulties encountered during evaluation of the
above complicated integral (\ref{in11}), we apply an asymptotic series
expansion method. As a result to series expansion, the integrand in (\ref{in11}) becomes
\begin{equation}
 \frac{2\omega \left[ \lambda mf+2\omega \left(
\lambda ^{2}+m^{2}r^{2}\right) \right] }{m^{2}r^{2}\left( \lambda
^{2}+m^{2}r^{2}\right) f^{2}}\simeq  \frac{4\omega ^{2}}{m^{4}r^{4}}+\frac{16\omega
^{2}M}{m^{4}r^{5}} .\label{in12}
\end{equation}%
\begin{equation*}+\frac{2\omega \left( \lambda m-2\omega \left( \lambda ^{2}+2\left(
Q^{2}-6M^{2}\right) m^{2}\right) \right) }{m^{6}r^{6}}
+\frac{2\omega M\left( \lambda m-4\omega \left( \lambda
^{2}+\left( 3Q^{2}-8M^{2}\right) m^{2}\right) \right) }{m^{6}r^{7}}+....
\end{equation*}%
Integrating (\ref{in12}) (for the similar procedure, the reader is refereed to [48]) and combining the result of the first integral (\ref{in32}), we obtain the following semi-analytical bound 
\begin{equation}
\sigma _{l}\left( w\right) \geq \sec h^{2}\left[ \frac{1}{2w}-\frac{\lambda 
}{2w}\left( \frac{4\omega ^{2}}{3\lambda ^{2}m^{4}r_{h}^{3}}+\frac{4\omega
^{2}M}{\lambda ^{2}m^{4}r_{h}^{4}}\right. \right.
\end{equation}%
\begin{equation*}
+\frac{2\omega \left( \lambda m-2\omega \left( \lambda ^{2}+2\left(
Q^{2}-6M^{2}\right) m^{2}\right) \right) }{5\lambda ^{2}m^{6}r_{h}^{5}}
\end{equation*}%
\begin{equation*}
\left. \left. +\frac{2\omega M\left( \lambda m-4\omega \left( \lambda
^{2}+\left( 3Q^{2}-8M^{2}\right) m^{2}\right) \right) }{3\lambda
^{2}m^{6}r_{h}^{6}}\right) \right] .
\end{equation*}%
The obtained transmission probability mainly depends on the mass of the Dirac field, BH mass and the magnetic charge. The potential and the behaviour of the GF in EYM BH for a
massive spin- 1/2 fermions are shown in Fig. 2. It is clear that the specific form of the GF is determined by potential shapes as in quantum
mechanics. We analyze that the potential gets lower when $Q$ increases
as shown in the left panel in Fig. 2, indicating that the GF increases as $Q$ decreases. Further, the bound $\sigma _{l}\left( w\right) $ exhibits the convergent behaviour and coming closer to $1$.

\subsubsection{Massless case}

In order to compute the GF for massless spin- 1/2 fermions emission, we use
the potential derived in (\ref{p12}). Thus, Eq. (\ref{gb1}) becomes%
\begin{equation}
\sigma _{l}\left( w\right) \geq \sec h^{2}\left( \frac{\lambda }{2\omega 
}\int_{r_{h}}^{+\infty }\frac{1}{r^{2}}\left( \lambda +\frac{rf^{\prime }}{2%
\sqrt{f}}-\sqrt{f}\right) dr\right) .  \label{in1}
\end{equation}%
To find the integral (\ref{in1}), we use the classical term-by-term integration technique to obtain asymptotic integral expansions, which requires the integrand to have a uniform asymptotic expansion in the integration variable [49]. Therefore the GF of spin- 1/2 fermions is%
\begin{equation}
\sigma _{l}\left( w\right) \geq \sec h^{2}\left[ \frac{\lambda }{2w}%
\left( \frac{M\left( 1-\lambda \right) }{r_{h}}+\frac{M\left( \lambda
-2\right) }{r_{h}^{2}}+\frac{\left( 5M^{2}+Q^{2}\right) \left( 5-2\lambda
\right) }{6r_{h}^{3}}\right. \right. .
\end{equation}%
\begin{equation*}
\left. \left. +\frac{\left( M^{3}-3MQ^{2}\right) }{4r_{h}^{4}}\right) \right]
\end{equation*}%
\begin{figure}
    \centering
{{\includegraphics[width=7.5cm]{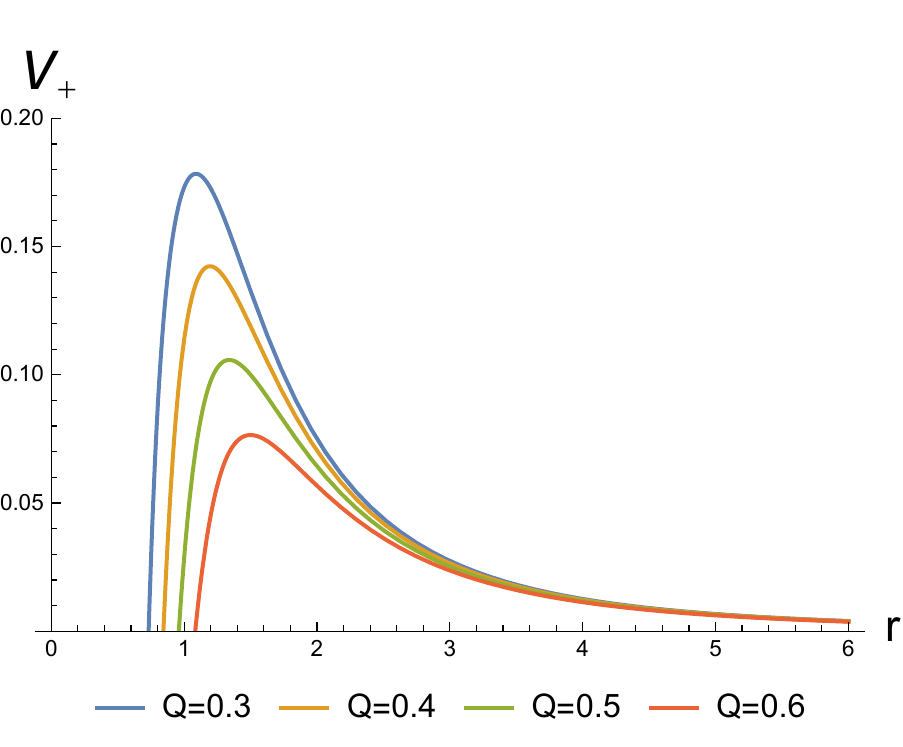} }}\qquad
    {{\includegraphics[width=7.5cm]{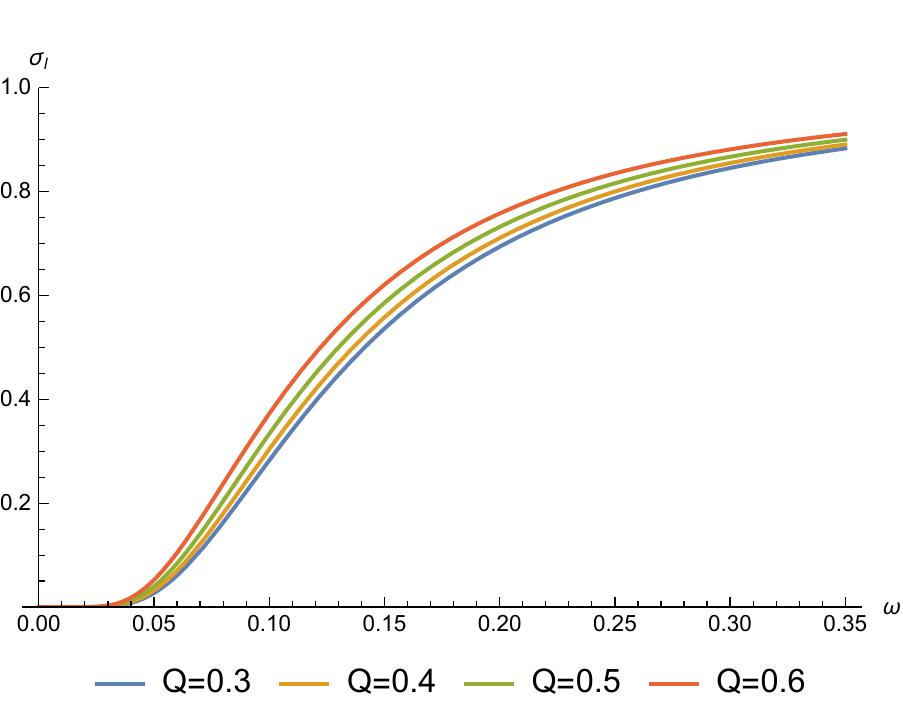}}}
    \caption{The potentials (\ref{p12})
for massless spin- 1/2 fermions for several values of $Q$ (left panel). The corresponding GF (right panel).
Here, we put the parameters as $\lambda =  1 = M  = \xi$.}%
    \label{fig1}%
\end{figure}
It is seen that the GF of massless spin- $1/2$ fermions increases as the value of magnetic charge parameter increases as shown in Fig. 3. This is comparable to the case of massive one. 

\subsection{Spin- 3/2 fermions emission}

To obtain the GF for spin- 3/2 fermions emission, we consider the potential (\ref{p32}). Then, Eq. (\ref{gb1}) becomes%
\begin{equation}
\sigma _{l}\left( w\right) \geq \sec h^{2}\left[ \frac{\lambda }{2\omega }%
\int_{r_{h}}^{+\infty }\left( \frac{\overline{\lambda }\left( 1-\overline{%
\lambda }^{2}\right) }{r^{2}\left( f-\overline{\lambda }^{2}\right) ^{2}}%
\right. \right.  \label{in18}
\end{equation}%
\begin{equation*}
\left. \times \left( \overline{\lambda }\left( 1-\overline{\lambda }%
^{2}\right) +\frac{\left( rf^{\prime }-2f\right) \left( f-\overline{\lambda }%
^{2}\right) }{2\sqrt{f}}-r\sqrt{f}f\right) dr\right] . 
\end{equation*}%
Here, the asymptotic series expansion method is applied. Then, the integral (\ref{in18}) is evaluated and the GF of spin- 3/2 fermions is%
\begin{equation}
\sigma _{l}\left( w\right) \geq \sec h^{2}\left[ \frac{1}{2w}\left( 
\frac{\left( 1-\overline{\lambda }\right) \overline{\lambda }M}{\left( 1+%
\overline{\lambda }\right) r_{h}^{2}}-\frac{\left( 1\mp \overline{\lambda }%
\right) \overline{\lambda }}{r_{h}}\right. \right. 
\end{equation}%
\begin{equation*}
\left. -\frac{\overline{\lambda }\left( \left( 3\left( \overline{\lambda }%
-1\right) ^{2}\left( 3+\overline{\lambda }\right) \right) M^{2}+\left( 
\overline{\lambda }+\overline{\lambda }^{2}-3\overline{\lambda }%
^{3}-3\right) Q^{2}\right) }{6\left( \overline{\lambda }-1\right) \left( 1+%
\overline{\lambda }\right) ^{2}r_{h}^{3}}\right] .
\end{equation*}%
\begin{figure}
    \centering
{{\includegraphics[width=7.5cm]{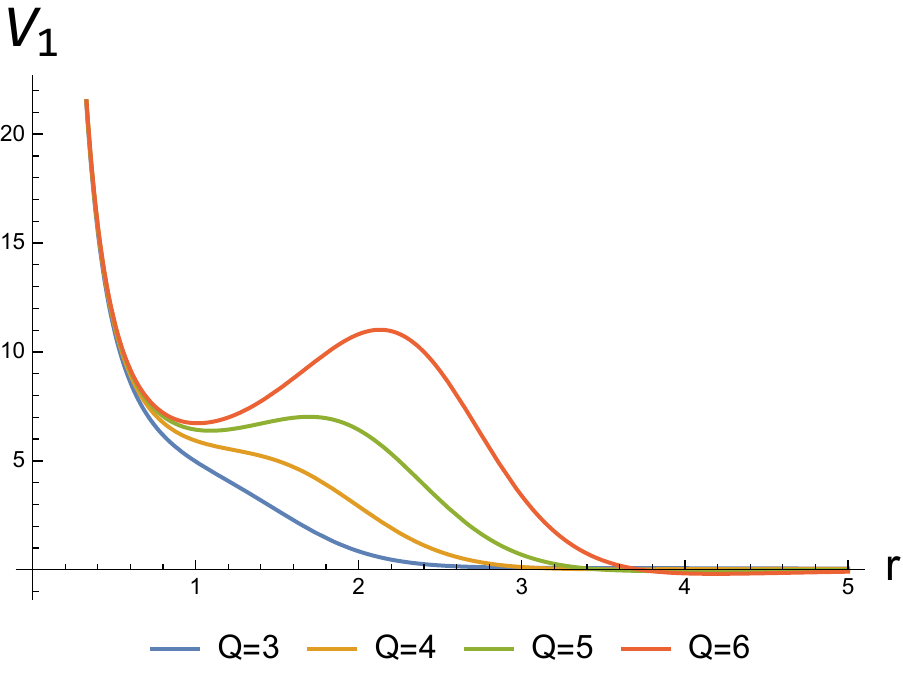} }}\qquad
    {{\includegraphics[width=7.5cm]{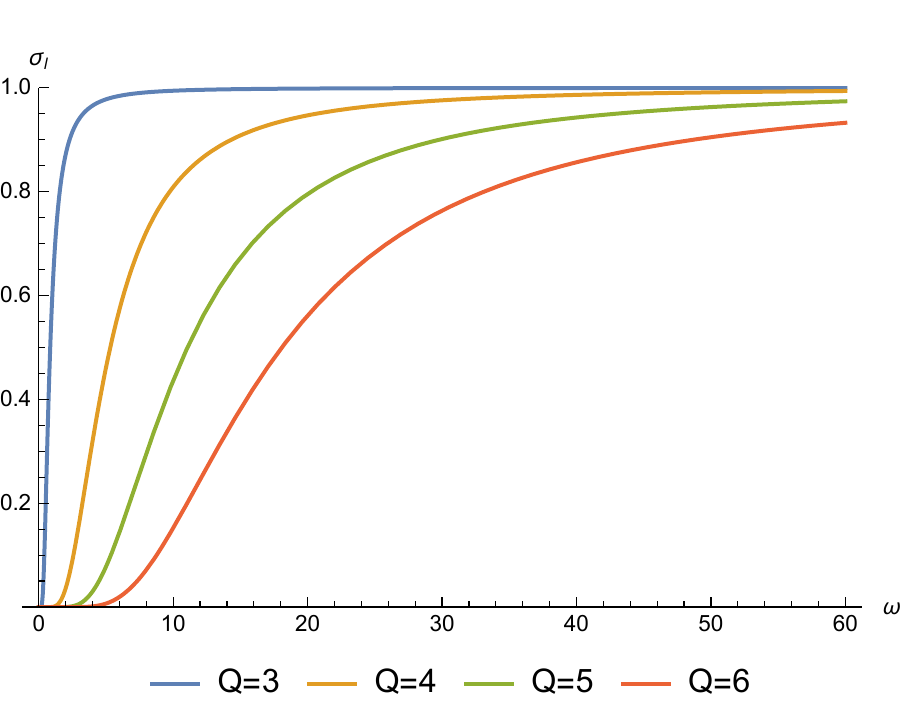}}}
    \caption{The potentials (\ref{p32})
for massless spin- 3/2 fermions for several values of $Q$ (left panel). The corresponding GF (right panel).
Here, we put the parameters as $\overline{\lambda} =  4, M = 1 = \xi$.}%
    \label{fig1}%
\end{figure}
Figure 4 shows the graphical behavior of the potential and the semi-analytic bounds for spin-3/2 fermions. We analyze that the higher the value of magnetic charge the lower the bound of GF as shown in Fig.4. The argument asserts that spin-3/2 fields have more self-interactions than spin-1/2 fields,  which will then make it more
difficult for the particles to pass through the potential.  

\subsection{Bosons emission}

The GFs of EYM\ BH via bosons emission is obtained by substituting the
effective potential derived in (\ref{v1}). Then Eq. (\ref{gb1}) becomes%
\begin{equation}
\sigma _{l} \left( \omega \right) \geq \sec h^{2}\left( \frac{1}{%
2\omega }\int_{r_{h}}^{+\infty }\left( \frac{l(l+1)}{r^{2}}+\frac{f^{\prime }%
}{r}\right) dr\right) .  \label{in21}
\end{equation}%
Although the integral (\ref{in21}) has an exact solution, it is
unfortunately not physical when $r\rightarrow \infty ,$ so we use the
asymptotic series expansion method instead. Therefore, the GF of bosons is%
\begin{equation}
\sigma _{l}\left( \omega \right) \geq \sec h^{2}\left[ \frac{1}{2\omega }%
\left( \frac{10M\zeta Q^{2}}{3r_{h}^{6}}+\frac{2Q^{2}}{3r_{h}^{3}}-\frac{M}{%
r_{h}^{2}}-\frac{l(l+1)}{r_{h}}\right) \right] .
\end{equation}%
Fig. 5 shows the graphical behaviors of the  potential and the semi-analytic bounds for bosons. It is shown that potential barriers enlarge with decreasing magnetic charge. The bounds of the GF reduce as a result. The situation is similar to that of spin- 1/2 fermions.
\begin{figure}
    \centering
{{\includegraphics[width=7.5cm]{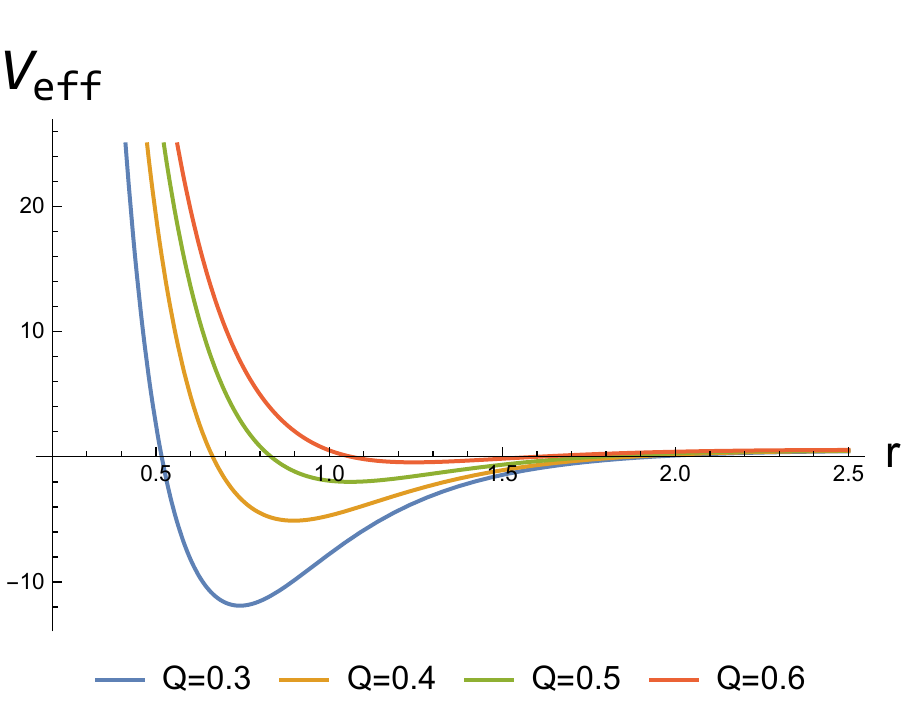} }}\qquad
    {{\includegraphics[width=7.5cm]{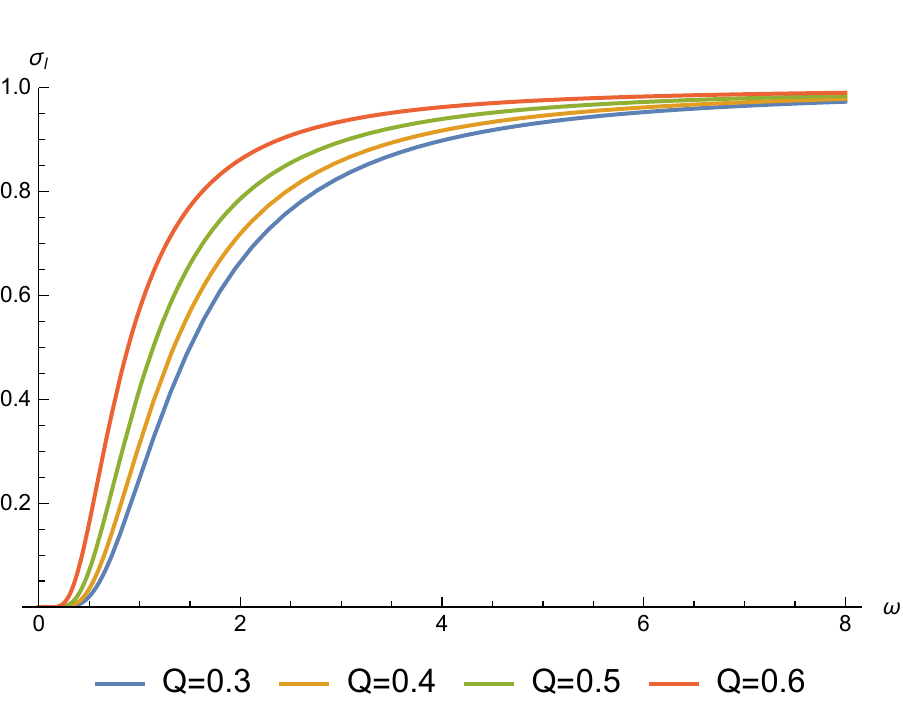}}}
    \caption{The potentials (\ref{v1})
for bosons with different for several values of $Q$ (left panel). The corresponding GF (right panel). Here, we put the parameters as $l =  1 = M = \xi$.}%
    \label{fig1}%
\end{figure}
\section{Conclusion}

In this paper, we studied the spinorial wave equations, including Dirac,
Rarita-Schwinger and Klein-Gordon equations, as well as the semi-analytical GFs in a regular non-minimal EYM BH spacetime. To summarize, we computed the GFs for fermions with
different spins and bosons by using semi-analytic method. With the regular
EYM BH metric, which uses the physical quantities mass of BH, non-minimal coupling parameters and magnetic
charge to characterize the spacetime around a magnetically charged
non-minimal magnetic BH, our goal is to determine  how the magnetic charge
parameter impacts the BH's thermal radiation.  We have used the Dirac, Rarita Schwinger, and Klein-Gordon equations of motion to describe spin- 1/2 fermions, spin- 3/2 and bosons, respectively. We divided the spinorial wave equations into radial and angular sets using the separation of variables method. Then, using tortoise coordinate the radial equations are transformed into Schr\"{o}dinger-like wave equation with their effective potentials.
In addition, we have analyzed graphically how magnetic charge variation affects Hawking radiation absorption potential. We then investigated the GFs for fermions with different spins and bosons in the regular EYM BH spacetime. The following are the main findings of this paper.

$\bullet $ We have found that the potentials for massive spin-1/2 fermions decrease and attenuate their peaks as the magnetic charge $Q$ rises (Fig. 2). As a result, as the
magnetic charge value increases, the GF rises.

$\bullet $ It is found that as $Q$ increases, the massless spin-1/2
fermion's potential height decreases (Fig. 3). We also found that the height
of the potential is greater for massive fermions than for massless ones. The
higher value of magnetic charge causes an increase in the GF for massless
spin-1/2 fermions.

$\bullet $ Our results showed that massless spin- 3/2 fermions have higher potential peaks as $Q$ increases (Fig. 4). GF is therefore inversely related to magnetic charge, as opposed to spin-1/2, whose GF decreases as $Q$ decreases. 

$\bullet $ We found that bosonic GF will increase as the magnetic charge
increases (Fig. 5) comparable to fermionic GFs with spin- 1/2.

$\bullet $ We noticed that all the bound $\sigma _{l}\left( w\right) $
approaches 1 as the value of $w$ rises.

In addition, it was demonstrated that the thermal emission of
Rarita-Schwinger fermions from a regular non-minimal EYM BH separates
fermions with different spins into distinct thermal radiations. We therefore
presented some semi-analytical results with future data that can be compared.
Finally, our results are then reduced to Schwarzschild BH when $Q = 0$ and Reisner-N\"{o}rdstrom BH when $\xi = 0$. We would be
interested in finding GF for the rotating regular magnetic BH solution of
the EYM theory [40] to uncover the effect of rotation. It is left for future investigation.\bigskip
\newline
{\LARGE Acknowledgements}\newline
The author would like to thank the Editor and anonymous Referee for their constructive suggestions and comments.\bigskip \newline
\bigskip
{\Large Data Availability}
\newline No data availability in this manuscript. 

\bigskip 

{\Large References\bigskip }

[1] The EHT Collaboration, Astrophys. J. 875, L1 (2019).

[2] The EHT Collaboration, Astrophys. J. Lett. 910, L13 (2021).

[3] K. Akiyama et al. [Event Horizon Telescope], Astrophys. J. Lett. 930,
L12 (2022).

[4] K. Akiyama et al. [Event Horizon Telescope], Astrophys. J. Lett. 930,
L13 (2022).

[5] B.P. Abbott et al. [LIGO Scientific and Virgo Collaborations], Phys.
Rev. Lett. 116(6), 061102 (2016)

[6] K. Akiyama et al. (Event Horizon Telescope Collaboration), Astrophys. J.
875, L1 (2019).

[7] K. Akiyama et al. (Event Horizon Telescope Collaboration), Astrophys. J.
875, L6 (2019).

[8] J. M. Bardeen, B. Carter, S.W. Hawking, Commun. Math. Phys. 31 161
(1973).

[9] S.W. Hawking, Commun. Math. Phys. 43, 199 (1975).

[10] S. W. Hawking, Phys. Rev. D 13 191(1976).

[11] S. Fernando, Gen. Relativ. Gravit. 37, 461 (2005).

[12] Jorge Escobedo, \textquotedblleft Greybody factors,\textquotedblright\
Master's Thesis, Uni. of Amsterdam 6 (2008).

[13] W. Kim, J.J. Oh, JKPS 52, 986 (2008).

[14] T. Harmark, J. Natario, R. Schiappa, Adv. Theor. Math. Phys. 14, 727
(2010).

[15] P. Boonserm, C.\thinspace H. Chen, T. Ngampitipan, and P. Wongjun,
Phys. Rev. D 104, 084054 (2021).

[16] K. Jusufi, M. Amir, M. Sabir Ali, S.D. Maharaj, Phys. Rev. D 102,
064020 (2020).

[17] M.K. Parikh, F. Wilczek, Phys. Rev. Lett. 85, 5049 (2000).

[18] P. Boonserm, T. Ngampitipan, P. Wongjun, Eur. Phys. J. C 79, 330 (2019).

[19] I. Sakalli, Phys. Rev. D 94, 084040 (2016).

[20] A. Al-Badawi, I. Sakalli, S. Kanzi, Ann. Phys. 412, 168026 (2020).

[21] A. Al-Badawi, S. Kanzi, I. Sakalli, Eur. Phys. J. Plus 135, 219 (2020).

[22] H. Gursel, I. Sakalli, Eur. Phys. J. C 80, 234 (2020).

[23] S. Kanzi, I. Sakall\i , Eur. Phys. J. C 81, 501 (2021).

[24] F. J. Zerill, Phys. Rev. D 9 860 (1974).

[25] S. S. Gubser, Commun. Math. Phys. 203 325 (1999).

[26] M. Cvetic, H. Lu, C. N. Pope, T. A. Tran,\ Phys. Rev. D 59 126002
(1999).

[27] I. Sakall\i , Phys. Rev. D 94 084040 (2016).

[28] S. S. Gubser, Physical Review D 56 4984 (1997).

[29] I. Sakalli and S. Kanzi, Turk. J. Phys. 46, no.2, 51-103 (2022).

[30] J. A. Smoller, A. G. Wasserman, and S. T. Yau, Commun. Math. Phys. 154,
377 (1993).

[31] P. Breitenlohner, D. Maison, D.H. Tchrakian, Class. Quantum Gravity 22
5201--5222 (2005).

[32] Y. Brihaye, E. Radu, D.H. Tchrakian, Phys. Rev. D 75 2 024022 (2007).

[33] S.H. Mazharimousavi, M. Halilsoy, Phys. Lett. B 659 471--475 (2008).

[34] D.O. Devecioglu, Phys. Rev. D 89 12 124020 (2014).

[35] A.B. Balakin, J.P.S. Lemos, Class. Quantum Gravity 22 1867--1880,
(2005).

[36] A.B. Balakin, S.V. Sushkov, A.E. Zayats, Rev. D 75 8 084042 (2007).

[37] A.B. Balakin, A.E. Zayats, Phys. Lett. B 644 294--298, (2007).

[38] A.B. Balakin, J.P.S. Lemos, A.E. Zayats, Phys. Rev. D 93 2 024008
(2016).

[39] Feng-Yuan Liu, Yi-Fan Mai, Wen-Yu Wu, Yi Xie, Physics Letters B 795
475--481 (2019).

[40] K. Jusufi, M. Azreg-A\"{\i}nou, M. Jamil, S. Wei, Q. Wu, and A. Wang,
Phys. Rev. D 103 024013 (2021).

[41] S. Devi, R. Roy and S. Chakrabarti, Eur. Phys. J. C 80, no.8, 760 (2020).

[42] S. B. Giddings, Phil. Trans. Roy. Soc. Lond. A 377, 20190029 (2019).

[43] J. Rayimbaev, B. Narzilloev, A. Abdujabbarov, B. Ahmedov, Galaxies 9,
71 (2021).

[44] P. Wongjun, C. H. Chen and R. Nakarachinda, Phys. Rev. D 101, 124033
(2020).

[45] C. H. Chen, H. T. Cho, A. S. Cornell, G. Harmsen and X. Ngcobo, Phys.
Rev. D 97 no.2 024038 (2018).

[46] C. H. Chen, H. T. Cho, A. S. Cornell and G. Harmsen, Phys. Rev. D 94,
no.4, 044052 (2016).

[47] M. Visser, Phys. Rev. A 59, 427-438 (1999).

\textcolor{blue}{[48] F. Faure, T. Weich, Commun. Math. Phys. (2015).}

\textcolor{blue}{[49] J. L´opez, J. Comput. Appl. Math. 102, 181 (1999).}

\end{document}